# A Coefficient of Determination ($R^2$) for Linear Mixed Models

Hans-Peter Piepho[1,*]

[1]Biostatistics Unit, Institute of Crop Science, University of Hohenheim, 70 593 Stuttgart, Germany

**email*: piepho@uni-hohenheim.de

SUMMARY. Extensions of linear models are very commonly used in the analysis of biological data. Whereas goodness of fit measures such as the coefficient of determination ($R^2$) or the adjusted $R^2$ are well established for linear models, it is not obvious how such measures should be defined for generalized linear and mixed models. There are by now several proposals but no consensus has yet emerged as to the best unified approach in these settings. In particular, it is an open question how to best account for heteroscedasticity and for covariance among observations induced by random effects. This paper proposes a new approach that addresses this issue and is universally applicable. It is exemplified using three biological examples.

## 1. Introduction

The coefficient of determination, also denoted as $R^2$, is perhaps the most popular measure of goodness of fit for linear models (LM) (Draper and Smith, 1998). It is defined as the proportion of the corrected sum of squares that is explained by the model. LM can be extended in two important ways. The first is to allow for random effects, leading to linear mixed models (LMM) (Searle et al., 1992), and the second is to allow for other distributions than the normal, leading to generalized linear models (GLM) (Nelder and McCullagh, 1989). Both extensions can be combined, amounting to generalized linear mixed models (GLMM) (McCulloch et al., 2005; Lee et al., 2006; Stroup, 2013). Users of GLMM procedures have a keen interest in measuring goodness of fit in a similar fashion as available for LM via $R^2$. Several proposals have been made for $R^2$ measures, some targeting GLM (Zhang, 2017), others targeting LMM (Edwards et al., 2008; Liu et al., 2008; Demidenko et al., 2012), and yet others covering GLMM (Nagakawa and Schielzeth, 2013; Jaeger et al., 2017; Nakagawa et al., 2017; Stoffel et al., 2017). Some of the proposals are applicable only to somewhat specialized settings and specific linear predictors (e.g., with independent random effects with constant variance). Although a lot of progress has been made recently, there does not as yet seem to be a generally agreed procedure that is broadly applicable to any GLM, LMM, or GLMM, including models with complex variance-covariance structure such as those needed, e.g., with repeated measures (Jennrich and Schluchter, 1986) or spatial data (Zhang, 2002).

For an $R^2$ measure for GLMM to find broad usage and be appealing to data analysts, it needs to have a simple definition in terms of variance explained that is easy to understand and communicate. Moreover, such a measure should reduce to the usual $R^2$ in case of LM. In the presence of random effects, the measure should also account for any covariance and heterogeneity of variance among observations. The purpose of this paper is to propose such a measure and illustrate its properties and usage. The key new idea compared to previous proposals is that with correlated data the total variance is best assessed based on the variance of pairwise differences rather than the marginal variance of the observed data, which is commonly used. The paper is organized as follows. To set the stage, we briefly revisit the definition of $R^2$ for LM. In the next section an $R^2$ measure for LMM is proposed. This is subsequently extended to the variance explained by random effects and to GLMMs. Three examples are used to illustrate the method.

## 2. The coefficient of determination for LM

The LM is given by

$$y = X\beta + e,  \tag{1}$$

where $y$ is a response vector of length $n$, with fixed effects vector $\beta$ and associated design matrix $X$, and $e \sim N(0, V = I_n \sigma_e^2)$ is the residual error vector. It is assumed that the fixed effects $\beta$ comprise a common intercept. In order to assess the explanatory power of the remaining effects in $\beta$, all of these effects are dropped, yielding the null model

$$y = 1_n \phi + e,  \tag{2}$$

where $1_n$ is a vector of ones and $\phi$ is the intercept. The residual error under this null model is distributed as $e \sim N(0, V_0 = I_n \sigma_{e0}^2)$. The coefficient of determination can be defined as

$$\Omega_\beta = \frac{\Delta\theta(V, V_0)}{\theta(V_0)},  \tag{3}$$

where $\theta(V)$ quantifies the total variance implied by the variance-covariance structure $V$, and $\Delta\theta(V, V_0) = \theta(V_0) - \theta(V)$ is the variance explained by the effects added in the full model relative to the null model. For LM we use $\theta(V_0) = \sigma_{e0}^2$, $\theta(V) = \sigma_e^2$, and $\Delta\theta(V, V_0) = \sigma_{e0}^2 - \sigma_e^2$ and hence

$$\Omega_\beta = \frac{\sigma_{e0}^2 - \sigma_e^2}{\sigma_{e0}^2} = 1 - \frac{\sigma_e^2}{\sigma_{e0}^2}. \tag{4}$$

If the variance components in (4) are estimated by maximum likelihood (ML), we obtain the ordinary coefficient of determination, $R^2$, for LM. If the variances are estimated by residual maximum likelihood (REML), the adjusted $R^2$ results.

**3. Measuring total variance in LMM**

The LMM can be defined as (Searle et al., 1992)

$$y = X\beta + Zu + e, \tag{5}$$

where $y$, $\beta$ and $X$ are as defined for LM, $u$ is a random effects vector $u \sim N(0, G)$ with design matrix $Z$ and $e$ is a residual error vector $e \sim N(0, R)$. Random vectors $u$ and $e$ are assumed to be mutually independent such that $y \sim N(X\beta, V)$ with

$$V = ZGZ^T + R. \tag{6}$$

Our coefficient of determination for fixed effects in LMM will have the same form as the one used for LM (equation 4), now using the variance-covariance structure $V$ for the full model in equation (5) and $V_0$ for the null model, in which $X\beta$ in (5) is replaced by $1_n\phi$. The definition of $\theta(V)$ should allow for possible heterogeneity of variance, i.e., heterogeneity of the diagonal elements in $V$, and for covariance between observations, i.e., non-zero off-diagonal elements in $V$. Moreover, the definition should be such that the measure reduces to the common $R^2$ for LM when random effects are dropped and $R = I_n \sigma_e^2$. Finally, we require the measure to be additive, i.e.,

$$\theta(V_1 + V_2) = \theta(V_1) + \theta(V_2), \qquad (7)$$

because this allows components of explained variance to be decomposed in a natural way. Two measures of total variance are considered. The first is based on the marginal variance (*mv*). For an individual response variable $y_i$ this is given by

$$mv(y_i) = v_{ii}, \qquad (8)$$

where $v_{ij}$ is the *ij*-th element of $V$. This marginal variance may be averaged across observations to yield the average marginal variance (AMV)

$$\theta^{AMV}(V) = \frac{1}{n}\sum_{i=1}^{n} mv(y_i) = \frac{1}{n} trace(V). \qquad (9)$$

The trace of a variance-covariance matrix is a common measure of total variance in multivariate analysis. The major downside of this criterion is that it does not account for covariances $v_{ij}$ $(i \neq j)$ (Mustonen, 1997; Johnson and Wichern, 2002, p.139). There are several alternative measures of total variance commonly used in multivariate analysis that allow accounting for covariance, such as the determinant of $|V|$, also known as generalized variance in this context (Wilks, 1932; Goodman, 1968; Seber, 1984; Stanek, 1990; Mostonen, 1997), or its standardized form, given by the positive *n*-th root of the generalized variance, $\theta(V) = |V|^{1/n}$, which is denoted as standardized generalized variance (SenGupta, 1987) and reduces to $\sigma_e^2$ for LM. This criterion is a nonlinear function of the elements of $V$, however, and as such fails to meet the additivity requirement (7). Moreover, in degenerate cases with very large correlations the generalized variance may become zero (Mostonen, 1997; Johnson and Wichern, 2002, p.130). Also, when $V$ is diagonal, $|V|^{1/n}$ is the geometric mean of the diagonal elements, whereas the arithmetic mean, as obtained by $\theta^{AMV}(V)$, seems more intuitively appealing. In addition there are other related measures of total or "explained" variance used to define multivariate test statistics (Pillai's trace, Hotelling-Lawley trace, Wilk's lambda, Roy's largest root; see, e.g., Johnson and Wichern, 2002, p.139), but all of these are also nonlinear functions of the elements of the variance-covariance matrices involved. For these reasons, we do not consider such alternative measures.

Instead, we propose a second measure that is also a linear function of the elements of $V$. The main motivation for this new measure is to account for the covariance among observations in an easily interpretable way. Rather than focusing on the marginal variance, we may consider the variance of a difference between observations. Arguably, this provides a natural way to account for covariance. For two individual response variables $y_i$ and $y_j$ $(i \neq j)$, the "semivariance" (*sv*) is defined as half the variance of $y_i - y_j$:

$$sv(y_i, y_j) = \frac{1}{2}\text{var}(y_i - y_j) = \frac{1}{2}(v_{ii} + v_{jj}) - v_{ij}. \tag{10}$$

The semivariance (or semivariogram) is a measure of variance commonly used for spatially correlated data (Isaacs and Srivastava, 1991). If there is a positive covariance $v_{ij}$ $(i \neq j)$, the semivariance is reduced in comparison with the average of the two marginal variances $v_{ii}$ and $v_{jj}$. Conversely, if the covariance is negative, the semivariance is increased compared to the average marginal variance. This naturally accounts for the interplay between variances $v_{ii}$ and covariances $v_{ij}$ in governing the overall variance between observations. The average semivariance (ASV), averaged across all pairs of observations in the data vector $y$, is the second proposed measure of total variance:

$$\theta^{ASV}(V) = \frac{2}{n(n-1)} \sum_{i=1}^{n-1} \sum_{j>i+1}^{n} sv(y_i, y_j) = \frac{1}{n-1} trace(VP_\mu), \tag{11}$$

where $P_\phi = I_n - n^{-1} 1_n 1_n^T$. It is readily verified that $\theta^{AMV}(V) = \theta^{ASV}(V) = \sigma_e^2$ when $V = R = I_n \sigma_e^2$ (i.e., for an LM) as required.

**4. Total variance exemplified for a balanced one-way model with random group effects**

To illustrate and compare the properties of $\theta^{AMV}(V)$ and $\theta^{ASV}(V)$ and to see how covariance is accounted for by the latter, consider the one-way random-effects model for $a$ groups and $m$ observations per group, given by

$$y = 1_n \phi + (I_a \otimes 1_m) u + e, \tag{12}$$

where $u \sim N(0, I_a \sigma_u^2)$ represents $a$ random group effects, and the error $e \sim N(0, I_n \sigma_e^2)$ represents $n = am$ random deviations from the respective group means. The variance of $y$ is

$$V = I_a \otimes (J_m \sigma_u^2 + I_m \sigma_e^2). \tag{13}$$

Under this model there is a positive covariance between observations within the same group, whereas observations from different groups are independent. The pairwise correlation among observations within the same group, the intra-class correlation, equals $\rho = \sigma_u^2 / (\sigma_u^2 + \sigma_e^2)$ and approaches unity when $\sigma_u^2 / \sigma_e^2 \to \infty$. For this model we find $\theta^{AMV}(V) = \sigma_u^2 + \sigma_e^2$ and $\theta^{ASV}(V) = (am-1)^{-1}(a-1)m\sigma_u^2 + \sigma_e^2$. It may be argued that $\theta^{ASV}(V)$ is the more meaningful measure of total variance here because the marginal variance of $\sigma_u^2 + \sigma_e^2$ only occurs for observations between groups but not within groups, where the variance amounts to only $\sigma_e^2$. Among the $n(n-1)/2$ pairs of observations, $am(m-1)/2$ pairs occur within groups and hence have variance $sv(y_i, y_j) = \sigma_e^2$ between them. The other $a(a-1)m^2/2$ pairs occur between groups and have variance $sv(y_i, y_j) = \sigma_u^2 + \sigma_e^2$ between them, which is equal to $mv(y_i, y_j)$. Averaging these variances across all pairs yields $\theta^{ASV}(V)$, confirming that this is indeed a meaningful measure of total variance. Also note that in the extreme case of having only one group $(a=1)$, we still find $\theta^{AMV}(V) = \sigma_u^2 + \sigma_e^2$, although for all pairs of observations the only relevant variance component is $\sigma_e^2$, which is well reflected by the fact that $\theta^{ASV}(V) = \sigma_e^2$ in this limiting case. This example illustrates that $\theta^{ASV}(V)$ not only accounts for the covariances among observations but also for the design of the study, given here by the number of groups, $a$, and the group size, $m$. By contrast, $\theta^{AMV}(V)$ cannot account for this structure, as it is focussed entirely on the diagonal of $V$.

We note in passing that the standardized generalized variance $|V|^{1/n} = \left[(\sigma_e^2)^{m-1}(\sigma_e^2 + m\sigma_u^2)\right]^{1/m}$ (Searle et al., 1992, p. 443) is obviously nonlinear in the variance parameters, hence violating the additivity requirement (7), and does not have the same intuitive interpretation as $\theta^{AMV}(V)$ and $\theta^{ASV}(V)$ do.

**5. A coefficient of determination for random effects in LMM**

We can also define a coefficient of determination for random effects $u$ as

$$\Omega_u = \frac{\theta(ZGZ^T)}{\theta(V_0)}. \tag{14}$$

The variance explained jointly by the fixed and random effects can be assessed by

$$\Omega_{\beta u} = \Omega_\beta + \Omega_u = 1 - \frac{\theta(R)}{\theta(V_0)}. \tag{15}$$

Furthermore, on account of the additivity requirement for $\theta(V)$ as per equation (7), we can compute partial coefficients of determination (Edwards et al., 2008) by partitioning the random effects as $Zu = Z_1 u_1 + Z_2 u_2 + ...$ and the associated variance as $ZGZ^T = Z_1 G_1 Z_1^T + Z_2 G_2 Z_2^T + ...$ and then using $\theta(Z_i G_i Z_i^T)$ to assess the explained variance at the level of individual random effects $u_i$.

## 6. Extension to GLMM

A GLMM has the linear predictor

$$\eta = X\beta + Zu \tag{16}$$

with fixed and random effects as defined for LMM in equation (5). The observation vector $y$ is assumed to have conditional expectation

$$E(y|\eta) = \mu = g^{-1}(\eta), \tag{17}$$

where $g(.)$ is a link function. It may be assumed that the conditional distribution of $y$ given $\eta$ is from an exponential family, in which case the model can be estimated, e.g., by full ML using adaptive Gaussian quadrature (Pinheiro and Bates, 1995) or by the Laplace method (Wolfinger, 1993). An alternative is to only make an assumption about the variance function $\text{var}(y|\mu)$ and allow overdispersion relative to parametric distributions in the exponential family. Such models can be fitted by pseudo-likelihood (Wolfinger and O'Connell, 1993) or penalized quasi-likelihood (Breslow and Clayton, 1993). Another option to allow for overdispersion is to include a random unit effect among the random effects $u$ on the linear predictor scale.

The residual variance occurs on the observed scale ("R-side"), which is not the linear predictor scale except when an identity link is used, whereas the variance due to random effects occurs on the linear predictor scale ("G-side") (Stroup, 2013). This makes it difficult to assess the total variance. A further complication is that variance on the observed scale depends on the conditional mean via the linear predictor (16), which in turn depends on the random effects and thus the overall variance structure. When considering a null model with fixed effects reduced to $1_n \phi$, the conditional mean structure $\eta$, given the random effects $u$, is altered. In order to preserve as much as possible of that conditional mean structure so that residual variance on the observed scale is modelled properly, it is suggested here to generally add a random unit effect $f$ with zero mean and $\text{var}(f) = R_f$ in the linear predictor (Nakagawa and Schielzeth, 2013):

$$\eta = X\beta + Zu + f . \qquad (18)$$

The main purpose of the random unit effect $f$ is to capture any unexplained variance on the linear predictor scale. Also note that the random unit effect accounts for overdispersion, and overdispersion is clearly to be expected if important predictors are omitted from the fixed effects structure $X\beta$.

In order to preserve additivity, it is necessary to assess the total variance on the linear predictor scale. To this end, we further introduce an auxiliary random residual vector $h^T = (h_1, h_2, ...)$ and consider the extended linear predictor $\tilde{\eta} = \eta + h$ and the conditional variance $\text{var}(g^{-1}(\tilde{\eta}) | \eta)$, asking which variance-covariance structure $R_h = \text{var}(h)$ leads to $\text{var}(g^{-1}(\tilde{\eta}) | \eta) \approx \text{var}(y | \mu)$ (Foulley et al., 1987). Note that this auxiliary random effect $h$ is not to be confused with the random unit effect $f$ included in the linear predictor (18). The auxilliary residual variance $R_h$ is then used along with the unit variance $R_f$ to define a variance-covariance matrix

$$\widetilde{V} = ZGZ^T + \widetilde{R} \qquad (19)$$

with $\widetilde{R} = R_f + R_h$ on the linear predictor scale, which is used in place of $V$ to assess the total variance for GLMMs.

For the special case of a binomial distribution and a logit link, a logistic distribution may be assumed for $h_i$, which has variance $\text{var}(h_i) = \pi^2/3$. Similarly, with a probit link we may assume a standard normal distribution for $h_i$, which has variance $\text{var}(h_i) = 1$ (Keen and Engel, 1997). These results are exact, i.e., $\text{var}(g^{-1}(\widetilde{\eta})|\eta) = \text{var}(y|\mu)$. For other distributions and links, no such exact results are available, so we take recourse to an approximation based on a Taylor series expansion (Foulley et al., 1987; Nakagawa et al., 2017). It is assumed here that

$$\text{var}(y|\mu) = A_\mu^{1/2} R A_\mu^{1/2}, \qquad (20)$$

where $A_\mu$ is a diagonal matrix with diagonal elements equal to evaluations of the variance function at the mean $\mu$ and $R$ is an unknown matrix. This variance structure allows for over-dispersion, and estimation in case of overdispersion requires pseudo-likelihood methods to be used (Wolfinger and O'Connell, 1993). Note that the LMM is a special case of this model with identity link and $A_\mu = I$. For GLMMs with conditional error distributions in the exponential family we have $R = I_n$. It may be assumed that

$$\text{var}(h) = R_h = W_\mu^{1/2} R W_\mu^{1/2}, \qquad (21)$$

where $W_\mu$ is a diagonal matrix with functions of the mean $\mu$ on the diagonal. Expanding $g^{-1}(\tilde{\eta})$ in a Taylor series about the mean $\eta$ of the linear predictor, we find that to first order

$$\operatorname{var}(g^{-1}(\tilde{\eta}) \mid \eta) \approx D_{\eta^*} W_\mu^{1/2} R W_\mu^{1/2} D_{\eta^*} \,, \qquad (22)$$

where $D_{\eta^*} = \operatorname{diag}\left[\partial g^{-1}(\tilde{\eta})/\partial \tilde{\eta}\right]_{\tilde{\eta}=\eta}$. Comparing coefficients between (20) and (22) yields $A_\mu = D_{\tilde{\eta}} W_\mu D_{\tilde{\eta}}$ and hence $W_\mu = D_{\tilde{\eta}}^{-1} A_\mu D_{\tilde{\eta}}^{-1}$. This approach is an extension of a method proposed by Foulley et al. (1987) for Poisson data and by Bennewitz et al. (2014) for overdispersed binomial and Poisson data. For example, with overdispersed Poisson data with $\operatorname{var}(y \mid \mu) = \phi\mu$ and log-link, we find $\operatorname{var}(h_i) \approx \phi\mu_i^{-1}$. For overdispersed binomial data with probit link we have $\operatorname{var}(h_i) \approx \phi \dfrac{\pi_i(1-\pi_i)}{[\varphi(\eta_i)] m_i}$, where $m_i$ is the binomial sample size of the $i$-th observation, $\pi_i$ is the $i$-th binomial probability and $\varphi(.)$ is the standard normal probability density (Bennewitz et al., 2014).

**7. Examples**

**Example 1**: We consider the data by Potthoff and Roy (1964) from an orthodontic study with 11 girls and 16 boys. At ages 8, 10, 12 and 14 years, the distance (mm) from the center of the pituitary to the pterygomaxillary fissue was measured for each child. This dataset has also been used by other authors to illustrate their proposed $R^2$ measures (Zheng, 2000; Edwards et al., 2008; Orelien and Edwards, 2008). We here fit the same models as in Edwards et al. (2008), using their nomenclature to identify models so our results can be directly compared to their Table II. The fixed-effects parts of the model are denoted as Models I to III. Model I comprises just an intercept and a linear age effect, Model II has a gender-specific intercept in addition and Model III has an interaction of gender and linear age added compared to Model II. The data clearly has a repeated-measures variance-covariance structure that needs to be modelled. The variance-covariance structures considered by Edwards et al. (2008) are: 1 = random intercept for individuals, 2 = random intercept and slope with unstructured 2 × 2 variance-covariance matrix, and 3 = 2 with heterogeneity of residual variance between girls and boys. The residual effects in these models are assumed to be independent with the same variance at each time point. For comparison, we added the following variance-covariance structures: 0 = independent residual with constant variance (corresponding to an LM), 4 = unstructured residual variance-covariance, 5 = unstructured residual variance-covariance with heterogeneity between girls and boys. Results are shown in Table 1 for the Akaike Information Criterion (AIC) based on REML and estimates of $\Omega_\beta$ based on both REML and ML. Covariance model 3 fitted best according to AIC with all three fixed-effects models (I, II, and III). For all variance-covariance models, the estimates of $\Omega_\beta$ indicate that inclusion of a gender main effect is important, whereas the interaction of gender with linear time provides only marginal additional improvement. The estimates of $\Omega_\beta$ for both ML and REML are rather lower than the $R^2$ values reported by Edwards et al. (2008; Table II) based on Wald-type F-statistics with different approximations of the denominator degrees of freedom. Comparison with the estimates of $\Omega_\beta$ for covariance model 0 for ML and REML, corresponding to ordinary and adjusted $R^2$ for LM, respectively, suggests that the $R^2$-values in Edwards et al. (2008) may overstate the goodness of fit. Estimates of $\Omega_\beta^{ASV}$ tend to be slightly larger than those of $\Omega_\beta^{AMV}$, so accounting for the covariance among observations does make a difference, albeit not a very large one in this example.

- Table 1 about here -

**Example 2**: Nakagawa and Schielzeth (2013) consider data on beetle larvae sampled from 12 populations. Within each population, larvae were obtained from two microhabitats, subjected to two different dietary treatments, and distinguished as male and female. Sexed pupae were reared in containers, each holding eight animals from the same population. There are three responses: (i) body length (Gaussian distribution), frequency of two male colour morphs (binary distribution), and (iii) the number of eggs laid by each female (Poisson distribution). The models fitted by Nakagawa and Schielzeth (2013) had independent homoscedastic random effects for population and container and fixed effects for treatment, habitat. For body length, the model also comprised a fixed main effect for sex. These models were fitted here using REML for body length and the Laplace method (Wolfinger, 1993) for the morph data and egg counts. For morph frequency we assumed a logit link and estimated the coefficient of determination accounting for binary variance on the linear predictor scale by setting $\text{var}(h_i) = \pi^2/3$ and $\text{cov}(h_i, h_j) = 0$ $(i \neq j)$, the variance of the logistic distribution. We estimated the predicted Poisson mean $\mu_i = \exp(\eta_i)$ based on the linear predictor (18) and, following Foulley et al. (1987), set $\text{var}(h_i) = \mu_i^{-1}$ and $\text{cov}(h_i, h_j) = 0$ $(i \neq j)$. For both morph data and egg counts, we fitted independent homoscedastic random effect for units ($R_f = I_n \sigma_f^2$) on the linear predictor scale in order to capture any unexplained variance not taken up by the other random effects.

Variance estimates for all traits agree with those of Nakagawa and Schielzeth (2013) to three decimal places (Table 2). For body length, most of the variance is explained by random effects. The estimate of $\Omega_{\beta u}$ is relatively close to the conditional $R^2$ ($R^2_{GLMM(c)}$) of 74% reported by Nakagawa and Schielzeth (2013), and their marginal $R^2$ ($R^2_{GLMM(m)}$) of 39% is close to the estimate of $\Omega_\beta$. For the binary morph data, the variances for population and container are slightly larger for the full model than the model with fixed effects dropped, suggesting that fixed effects have no explanatory power for this trait. Consequently, $\Omega_\beta$ is estimated to be near zero. By way of comparison, Nakagawa and Schielzeth (2013) report a marginal $R^2$ for the fixed effects of 8%, which is somewhat unexpected given the smaller estimated variances under the null model. This likely results from the fact that they only use the fit of the full model to compute $R^2$, using the fixed-effects estimates themselves to estimate the variance explained by these. By contrast, our approach uses both the fits of the full and null models and assesses explained variance based on estimates of $V$ and $V_0$. The egg counts have larger estimated coefficients of determination $\Omega$ (Table 3), but again these are smaller than those reported in Nakagawa and Schielzeth (2013).

- Table 2 about here -

**Example 3**: Zhang (2017) used data from a study of nesting horseshoe crabs (Agresti, 1996) to illustrate different definitions of $R^2$ for GLM. A total of 173 crabs were assessed for colors (C), spine conditions (SC), carapace width (CW), and weight (W), each with a male crab attached to her in her nest. To investigate the effect of these factors on the number of satellites, i.e., any other males riding near a female crab, a GLM with log-link and Poisson distribution was fitted by Zhang (2017). Here, we fit a GLMM with independent homoscadastic random effect for units ( $R_f = I_n \sigma_f^2$ ) in the linear predictor so that the unexplained variance under the reduced models can be captured by that effect and thus allocated to the model's variance. The models are fitted by both the Laplace method (Wolfinger, 1993) and adaptive Gaussian quadrature (Pinheiro and Bates, 1995). The auxilliary variance was set to $\text{var}(h_i) = \mu_i^{-1}$ and $\text{cov}(h_i, h_j) = 0$ $(i \neq j)$ as in Example 2. We fitted the same fixed effects models as Zhang (2017), so our results may be directly compared to alternative $R^2$ measures in Table 1 of that paper. Because in this case $\widetilde{V}$ is diagonal, $\Omega_\beta^{ASV}$ and $\Omega_\beta^{AMV}$ coincide. On average, estimates of $\Omega_\beta$ in Table 3 are slightly larger than the $R^2$ measures proposed by Zhang (2017), but give a similar ranking among models.

We also fitted a GLM by pseudo-likelihood and residual pseudo-likelihood (Wolfinger and O'Connell, 1993), using a Taylor series expansion around (18) (Table 4). Coefficients of determination based on pseudo likelihood are slightly lower in magnitude compared to those in Table 3. Residual pseudo-likelihood (akin to REML) yields smaller coefficients of determination than pseudo-likelihood (akin to ML) as expected. The residual pseudo-likelihood method picks the model with the single covariate W as the best, whereas the other methods, all of which use ML, select more complex models. This outcome illustrates that use of REML-like methods for estimating variance components in GLMM leads to coefficients of determination that behave like adjusted $R^2$ for LM. Zhang (2017) also found the same best model using an adjusted $R^2$ measure for GLM.

- Tables 3 and 4 about here -

## 8. Discussion

There are several proposals of $R^2$ measures for GLM, LMM and GLMM in the literature, and most of them share the desirable property to reduce to the common $R^2$ in case of LM (Cameron and Windmeijer, 1996). This may be considered a necessary but not a sufficient condition, however. There are several proposals, mostly targeting GLM, that are functions of the maximized likelihood of quasi likelihood under the full and null models, most notable among them the likelihood ratio statistic (Maddala, 1983; Cox and Snell, 1983; Magee, 1990; Nagelkerke, 1991; Zheng, 2000) and the Kullback-Leibler divergence (Cameron and Windmeijer, 1997). While these measures certainly have their merits, they are more difficult to communicate to research scientists primarily familiar with ordinary least squares for LM and the concept of "explained variance".

There are several measures, mainly proposed for LMM, that make use of quadratic forms of $y$. For example, Buse (1973), Kramer (2005) and Demidenko et al. (2012) assess the unexplained variance for fixed effects based on the weighted residual sum of squares, $SS_W = (y - X\beta)^T V^{-1} (y - X\beta)$. While this measure does account for covariances among observations, it does so via $V^{-1}$ rather than the variance $V$, which may be difficult to explain to non-statisticians. Also, it is not immediately obvious how exactly $SS_W$ is related to the variance of the data $V$. Here, I am not refering to the mathematical relationship, which is obvious, but for an intuitive explanation that is easily grasped by a research scientist. Finally, the approach only works for LMM but not for GLMM.

Edwards et al. (2008) and Jaeger et al. (2017) proposed $R^2$ measures that are motivated by the fact that for LM the F-statistic for comparing the full and reduced models can be written as a function of $R^2$ and vice versa. This fact is used to define $R^2$ measures for LMM and GLMM by analogy based on the Wald-type F-statistic for the same type of comparison. A practical difficulty with this approach is that residual degrees of freedom need to be determined, and there are several approximations in use, leading to different values of $R^2$ [see Table II in Edwards et al. (2008) for a lucid example]. Also, Wald-type F-statistics involve $V^{-1}$, as does $SS_W$. A further complication is that different mixed model packages compute F-statistics differently, so with complex variance-covariance models there may be differences between packages (Piepho and Edmondson, 2018). Furthermore, whereas the $R^2$ for LM can indeed be written as a function of an F-statistic, that representation does not lend itself so well to communicate the interpretation in terms of explained variance. For that purpose it is preferable to express the $R^2$ as in equation (3) or in terms of sums of squares. But such an analogous representation does not seem to be forthcoming when a Wald-type F-statistic is used to define $R^2$ for LMM and GLMM.

Next, there are several proposals for $R^2$ measures that exclusively operate on unweighted sums of squares of deviations between observed data and fitted values (e.g., Vonesh et al., 1996; Liu et al., 2008) or the unweighted sum of squares of fitted values for $X\beta$ (Nakagawa and Schielzeth, 2013; Nakagawa et al., 2017). These measures do not make explicit use of $V$, other than in the generalized least squares estimator of $\beta$. Hence, it may be said that these measures do not account for covariances among observations or heterogeneity of variance.

This paper has proposed a new method to assess the coefficient of determination that is generally applicable to any GLM, LMM or GLMM, regardless of the variance-covariance structure used, and reduces to $R^2$ and adjusted $R^2$ for LM. The proposed coefficient of determination is assessed on the linear predictor scale, allowing an additive decomposition of total variance in case of random effects, which would not be forthcoming with approaches that target variance on the observed scale, such as the recent proposal by Zhang (2017) for GLM. When variance components are estimated by ML (approximate or exact), the coefficient of determination behaves like the ordinary $R^2$ for LM, whereas with REML-like methods of estimation (Piepho et al., 2018), the coefficient of determination works like the adjusted $R^2$ for LM. Two measures of total variance were considered that coincide for independent data. The first, $\theta^{AMV}(V)$, uses only the marginal variance, i.e., the diagonal elements of $V$. Similar approaches have been used by many authors proposing $R^2$ measures for LMM and GLMM, primarily when $G$ has a simple variance-components structure (Snijders and Bosker, 1999; Nakagawa and Schielzeth, 2013; Nakagawa et al., 2017). A major limitation of this measure, however, is that it cannot account for the covariance among observations. A key feature of the approach based on the second measure, $\theta^{ASV}(V)$, which is novel, is that covariance among observations is taken into account by assessing total variance in terms if the mean variance of a difference among observations. This idea has roots in geostatistics where semivariograms are used routinely to describe spatial variance and covariance (Isaacs and Srivastava, 1991), and it also bears some relation to experimental design, where efficiency may be assessed in terms of the average pairwise variance of a difference among treatments (Bueno Filho and Gilmour, 2003; Williams and Piepho, 2015), and to methods used for the estimation of heritability in plant breeding experiments. Note that the heritability has an interpretation of $R^2$ for the regression of phenotypes on genotypes (Falconer and Mackay, 1996). In fact, when $y$ is a random vector of adjusted genotype means based on an analysis of an individual experiment or of a series of experiments, $R$ is the associated variance-covariance matrix of adjusted means, $G = I_n \sigma_u^2$ is the genetic variance-covariance matrix, and $X\beta = 1_n \phi$, then $\Omega_u$ is equivalent to the broad-sense heritability defined in eq. (19) in (Piepho and Möhring, 2007).

All analyses were implemented using the MIXED and GLIMMIX procedures of SAS. These procedures were used to fit the full and null LMM, producing estimates of $V$, $V_0$ and $R$. For GLMM, output from GLIMMIX was post-processed to obtain estimates of $\tilde{V}$, $\tilde{V}_0$ and $\tilde{R}$. These outputs were then submitted to a macro that computes estimates of $\Omega_\beta$, $\Omega_u$ and $\Omega_{\beta u}$. The full code for the three examples is provided with the Supporting Information.

ACKNOWLEDGEMENTS

Hans-Peter Piepho was supported by the German Research Foundation (DFG; grant PI 377/18-1).

**Table 1**

Coefficients of determination ($\Omega_\beta^{ASV}, \Omega_\beta^{AMV}$) of LMM for repeated-measures dental data (Example 1).

| Variance-covariance model[§] | Fixed effects model[$] | AIC (REML) | REML $\Omega_\beta^{ASV}$ | REML $\Omega_\beta^{AMV}$ | ML $\Omega_\beta^{ASV}$ | ML $\Omega_\beta^{AMV}$ |
|---|---|---|---|---|---|---|
| 0 | I   | 511 | 0.249[&] | 0.249[&] | 0.256[#] | 0.256[#] |
|   | II  | 487 | 0.398[&] | 0.398[&] | 0.410[#] | 0.410[#] |
|   | III | 486 | 0.406[&] | 0.406[&] | 0.423[#] | 0.423[#] |
| 1 | I   | 451 | 0.254 | 0.249 | 0.262 | 0.257 |
|   | II  | 442 | 0.391 | 0.388 | 0.412 | 0.410 |
|   | III | 438 | 0.402 | 0.399 | 0.426 | 0.423 |
| 2 | I   | 451 | 0.353 | 0.349 | 0.363 | 0.360 |
|   | II  | 443 | 0.471 | 0.469 | 0.492 | 0.491 |
|   | III | 441 | 0.481 | 0.479 | 0.505 | 0.503 |
| 3 | I   | 432 (best) | 0.330 | 0.327 | 0.339 | 0.336 |
|   | II  | 425 (best) | 0.448 | 0.446 | 0.468 | 0.467 |
|   | III | 422 (best) | 0.463 | 0.461 | 0.486 | 0.484 |
| 4 | I   | 455 | 0.383 | 0.381 | 0.393 | 0.391 |
|   | II  | 449 | 0.495 | 0.494 | 0.516 | 0.515 |
|   | III | 445 | 0.506 | 0.505 | 0.528 | 0.527 |
| 5 | I   | 452 | 0.239 | 0.234 | 0.241 | 0.235 |
|   | II  | 446 | 0.385 | 0.382 | 0.406 | 0.403 |
|   | III | 441 | 0.401 | 0.398 | 0.429 | 0.426 |

[§] 0 = independent residual with constant variance (corresponding to an LM), 1 = random intercept for individuals, 2 = random intercept and slope with unstructured 2 × 2 variance-covariance matrix, 3 = 2 with heterogeneity of residual variance between girls and boys, 4 = unstructured residual variance-covariance, 5 = 4 = unstructured residual variance-covariance with heterogeneity between girls and boys.

[$] I = linear age effect only, II = linear age effect and gender main effect, III = II with linear age-by-gender interaction added

[&] Equivalent to $R^2$ for LM

[#] Equivalent to adjusted $R^2$ for LM

**Table 2**

Coefficients of determination ($\Omega_\beta, \Omega_u, \Omega_{\beta u}$) and variance component estimates (obtained by Laplace method) for beetle data (Example 2).

| Parameter | Traits | | | | | |
| --- | --- | --- | --- | --- | --- | --- |
| | Body length (Gaussian)[$] | | Morph data (binary)[$] | | Egg count (Poisson)[$] | |
| Variance components[&] | Null model[§] | Full model | Null model[§] | Full model | Null model[§] | Full model |
| Population | 1.181 | 1.379 | 0.946 | 1.110 | 0.303 | 0.304 |
| Container | 2.206 | 0.235 | 0.000 | 0.006 | 0.012 | 0.023 |
| Units | 1.224 | 1.197 | 0.000 | 0.000 | 0.171 | 0.100 |
| Coefficient of determination | $\Omega^{ASV}$ | $\Omega^{AMV}$ | $\Omega^{ASV}$ | $\Omega^{AMV}$ | $\Omega^{ASV}$ | $\Omega^{AMV}$ |
| $\Omega_\beta$ | 0.4009 | 0.3906 | −0.0377 | −0.0402 | 0.0092 | 0.0091 |
| $\Omega_u$ | 0.3330 | 0.3498 | 0.2467 | 0.2635 | 0.0474 | 0.0512 |
| $\Omega_{\beta u}$ | 0.7339 | 0.7405 | 0.2089 | 0.2233 | 0.0566 | 0.0603 |

[§] All fixed effects dropped from full model.
[&] All estimates agree closely with those reported in Nakagawa and Schielzeth (2013) up to the third decimal place.
[$] An LMM was fitted to the Gaussian data, whereas GLMM were fitted to the binary and Poisson data with logit-link and log-link, respectively. For the binary model, the auxilliary variance was set to $\text{var}(h_i) = \pi^2/3$ and $\text{cov}(h_i, h_j) = 0$ $(i \neq j)$. For the Poisson model the auxilliary variance was set to $\text{var}(h_i) = \mu_i^{-1}$ and $\text{cov}(h_i, h_j) = 0$ $(i \neq j)$.

**Table 3**

Coefficient of determination $\Omega_\beta$ and estimates of unit variance $\sigma_f^2$ component for GLMM fitted to crab count data (Poisson GLMM with log-link) by the Laplace method and by Gaussian quadrature (Example 3).

| Covariate model[$] | Estimation method | | | |
|---|---|---|---|---|
| | Laplace | | Gaussian quadrature | |
| | $\Omega_\beta^{ASV} = \Omega_\beta^{AMV}$ [§] | Unit variance | $\Omega_\beta^{ASV} = \Omega_\beta^{AMV}$ [§] | Unit variance |
| (C,SC,CW,W) | 0.203 | 0.96 | 0.212 | 0.98 |
| (C,CW,W) | 0.201 | 0.96 | 0.210 | 0.99 |
| (C,CW) | 0.181 | 0.99 | 0.190 | 1.01 |
| (C,W) | 0.206 | 0.96 | 0.213 | 0.98 |
| (CW,W) | 0.184 | 0.98 | 0.192 | 1.01 |
| C | 0.055 | 1.14 | 0.059 | 1.18 |
| SC | 0.032 | 1.17 | 0.032 | 1.21 |
| CW | 0.162 | 1.01 | 0.170 | 1.04 |
| W | 0.186 | 0.98 | 0.193 | 1.01 |
| – | – | 1.21 | – | 1.25 |

[§] Total variance was assessed on the linear predictor scale based on $\widetilde{R}$. The linear predictor had a random unit effect to capture unexplained variance. The auxilliary variance was set to $\text{var}(h_i) = \mu_i^{-1}$ and $\text{cov}(h_i, h_j) = 0$ $(i \neq j)$.

[$] C = colour, SC = spine condition, CW = carapace width, W = weight

**Table 4**

Coefficient of determination $\Omega_\beta$ and estimates of unit variance $\sigma_f^2$ for GLMM fitted to crab count data (Poisson GLMM with log-link) by the (residual) pseudo-likelihood method (Example 3).

| Covariate model$ | Estimation method † | | | |
|---|---|---|---|---|
| | Residual pseudo-likelihood‡ | | Pseudo-likelihood& | |
| | $\Omega_\beta^{ASV} = \Omega_\beta^{AMV}$ § | Unit variance | $\Omega_\beta^{ASV} = \Omega_\beta^{AMV}$ § | Unit variance |
| (C,SC,CW,W) | 0.097 | 0.83 | 0.175 | 0.75 |
| (C,CW,W) | 0.118 | 0.81 | 0.173 | 0.75 |
| (C,CW) | 0.111 | 0.82 | 0.156 | 0.77 |
| (C,W) | 0.133 | 0.80 | 0.177 | 0.75 |
| (CW,W) | 0.136 | 0.79 | 0.158 | 0.76 |
| C | 0.010 | 0.91 | 0.046 | 0.86 |
| SC | 0.003 | 0.92 | 0.027 | 0.88 |
| CW | 0.128 | 0.80 | 0.139 | 0.78 |
| W | 0.148 | 0.78 | 0.160 | 0.76 |
| – | – | 0.92 | – | 0.91 |

§ Total variance was assessed on the linear predictor scale based on $\widetilde{R}$. The linear predictor had a random unit effect to capture unexplained variance. The auxilliary variance was set to $\text{var}(h_i) = \mu_i^{-1}$ and $\text{cov}(h_i, h_j) = 0$ $(i \neq j)$.

$ C = colour, SC = spine condition, CW = carapace width, W = weight
† Taylor series expansion around $\eta = X\beta + Zu + f$
‡ Option RSPL in GLIMMIX procedure of SAS
& Option MSPL in GLIMMIX procedure of SAS